\documentclass[a4paper,11pt]{article}
\usepackage{longtable}
\usepackage[fleqn]{amsmath}
\usepackage{amssymb,amsthm}
\usepackage{graphicx}
\usepackage{bm}
\usepackage{footmisc}
\usepackage{afterpage}
\usepackage{float}
\usepackage{lscape}

\usepackage{longtable}

\usepackage[square,comma,sort&compress,numbers]{natbib}

\usepackage[dvipsnames]{xcolor}
\definecolor{gray75}{gray}{0.4}

\usepackage{fullpage}
\usepackage{etex}
\usepackage[utf8]{inputenc}
\usepackage[T1]{fontenc}
\usepackage{lmodern}
\usepackage[nolist,nohyperlinks]{acronym}
\usepackage{booktabs}
\usepackage{hepunits}
\usepackage{units}

\usepackage{multirow}

\usepackage{epigraph}
\epigraphsize{\small\itshape}
\usepackage{csquotes}

\usepackage{slashed}
\usepackage[section]{easy-todo}
\usepackage{caption}
\usepackage{subcaption}
\usepackage{setspace}
\usepackage{notoccite}
\usepackage{listings}
\usepackage{enumerate}

\usepackage[hyperfootnotes=false,pdfpagelabels]{hyperref}  

\hypersetup{%
    colorlinks=true, linktocpage=true, pdfstartpage=3, pdfstartview=FitV,%
    breaklinks=true, pdfpagemode=UseNone, pageanchor=true, pdfpagemode=UseOutlines,%
    plainpages=false, bookmarksnumbered, bookmarksopen=true, bookmarksopenlevel=1,%
    hypertexnames=true, pdfhighlight=/O,
   urlcolor=gray75, linkcolor=gray75, citecolor=gray75,
     pdftitle={Non-perturbative phenomenology},%
     pdfauthor={Leonardo Antunes Pedro},%
}

\usepackage{mciteplus}

\newcommand*{\bea}{\begin{eqnarray}}
\newcommand*{\eea}{\end{eqnarray}}
\newcommand*{\be}{\begin{equation}}
\newcommand*{\ee}{\end{equation}}

\newcommand*{\tr}{\mathrm{tr}}

\newcommand{\bma}{\begin{pmatrix}}
\newcommand{\ema}{\end{pmatrix}}

\graphicspath{{./Images/}}

\renewcommand{\textflush}{flushepinormal}

\theoremstyle{plain}
\newtheorem{thm}{Theorem}

\newtheorem*{rmk*}{Note}

\theoremstyle{definition}
\newtheorem{defn}[thm]{Definition}
\newtheorem*{defn*}{Definition}

\theoremstyle{remark}

\makeatletter
\renewcommand{\@epitext}[1]{
\itshape \begin{minipage}{\epigraphwidth}\begin{\textflush} #1
\end{\textflush}\end{minipage}\vspace{1ex}}

\makeatother
\title{Relating spontaneous and explicit symmetry breaking in the presence of the Higgs mechanism}

\author{Leonardo Pedro\\
Centro de F\'isica Te\'orica de Part\'iculas, Universidade de Lisboa,\\
Av. Rovisco Pais, P-1049-001 Lisboa, Portugal}

\date{\today}

\begin{document}

\maketitle

\begin{abstract}
  One common way to define spontaneous symmetry breaking involves explicit symmetry breaking. This definition can be used in any approach to Effective Field Theory,
from perturbation theory to lattice simulations. It allows us to study the spontaneous breakdown of global symmetries without assuming that
the local gauge symmetry is spontaneously broken. This is important since perturbation theory is insufficient to study extended Higgs sectors:
it is insufficient to predict the physical spectrum of the $SU(5)$ Grand Unified Theory (Georgi-Glashow) or to predict the spontaneous breakdown of global symmetries.
 
We also study background symmetries: these are symmetries that despite they are already explicitly broken, can be still spontaneously broken.
We analyse examples where a background CP (charge-parity) symmetry is not spontaneously broken: in the Standard Model, in rephasing symmetries and in geometrical CP-violation. 

We show that all fields are real representations of the group of symmetries, since CP is a unitary transformation.
There are consequences: to study accidental symmetries (e.g. custodial symmetry, pseudo-golstone bosons) we must consider real representations;
CP is a symmetry of order 4 if the neutrinos are Majorana particles and
the notion of CP-violating phases is inconsistent in some Lagrangians;
a recent claim that a toy model exhibits physical CP-violation while the CP symmetry is conserved by the Lagrangian and the vacuum is false.
\end{abstract}

\section{Introduction}


There are several definitions of spontaneous breaking of global symmetries~\cite{yangising,symmetrybreaking,*symmetries}, all are related with the phase transitions of a system\footnote{The converse is not true: there are topological phase transitions which do not involve symmetry breaking~\cite{topologicalnobel}.}. In one common definition\footnote{In statistical mechanics~\cite{yangising}; in the Standard Model~\cite{Englert:2004yk} and in two-Higgs-doublet models (in the context of lattice simulations~\cite{Lewis:2010ps,Maas:2016qpu})}, spontaneous symmetry breaking is a particular case of explicit symmetry breaking.

Let $\phi$ be a field, let $g$ be a global transformation acting on the field $\phi\to g(\phi)$. 

The expectation value of the field is $<\phi>_{J,N}$.
It depends on an external field $J$ which is not $g$-invariant (i.e. $J\neq g(J)$), while
$N$ is the size of the system. In the absence of the external field the system is $g$-invariant, thus $<(\phi-g(\phi))>_{J=0,N}=0$.

For finite size $N$, we assume that the system is well behaved with continuous expectation values\footnote{It is not strictly required that the expectation values are continuous for finite $N$ to have spontaneous symmetry breaking~\cite{symmetrybreaking,*symmetries}, but the systems with local interactions (e.g. the Ising model or gauge theories) share this property.
There are also systems where the thermodynamic limit makes no sense~\cite{smallsystems}, requiring a more general definition of phase transition.} as a function of $J$,
that is $\lim_{J\to 0}<(\phi-g(\phi))>_{J,N}=0$.

\begin{defn}[In statistical mechanics]
\label{stat}
There is spontaneous symmetry breaking when the expectation value is finite and breaks the symmetry generated by $g$,
for an infinite size $N$ and an arbitrarily small external field $J$, i.e.
\begin{align*}
\lim_{J\to 0}\{\lim_{N\to \infty}<(\phi-g(\phi))>_{J,N}\} \neq 0
\end{align*}
\end{defn}

The limits may not commute, because the (pointwise) limit of a convergent sequence of continuous functions is not necessarily continuous. 

Other definitions in the context of statistical mechanics
do not involve explicit symmetry breaking\footnote{Such definitions are not based on the existence of expectation values that explicitly break the symmetry, since that would not be possible by definition of the system's expectation value with $J=0$.  We prefer Def.~\ref{stat} because it allows us to study the symmetries of the system at the Lagrangian level, independently of the particular Quantum Field Theory framework (e.g. perturbative/continuum or non-perturbative/UV-cutoff, scattering processes or bound-states), as we want to use several frameworks for phenomenology studies.}, and are based instead on: a long-range order parameter which is the expectation value of a $G$-symmetric function $f(A)$ (e.g. the modulus $f(A)=|A|$) of an operator $A$ which breaks $G$ and it is invariant under translations in space-time; or a conditional expectation value of some operator $A$ given some condition $C=0$ that breaks the symmetry; or a two-point correlation function with the points at an infinite distance from each other (related with boundary conditions)~\cite{yangising}. It is widely accepted that these definitions should be all equivalent to Def.~\ref{stat} (\cite[See Sec. 10.C]{symmetrybreaking} also in the Ising model~\cite{yangising}), although it is not easy to prove it because the systems with or without external field are physically different~\cite{exactly}.

When it comes to quantum non-abelian gauge field theories, the theories themselves lack a non-perturbative mathematical definition~\cite{prize}, so it is even more difficult to relate these different definitions.
By analogy with statistical mechanics\footnote{The correlation functions of quantum field theory can be defined as the Wick-rotation of correlation functions of a statistical field theory~\cite{nonperturbativefoundations}.}, these definitions make sense within the framework of quantum phase transitions~\cite{qpt,meanfield,wignerspontaneous} (even at zero temperature). 
In the presence of the Higgs mechanism, there is yet another definition of spontaneous symmetry breaking, most common in the context of perturbation theory of the Electroweak interactions:

\begin{defn}[Electroweak symmetry breaking]
\label{pt}
The vacuum expectation value (vev) of the Higgs field is determined  by one of the possible minima of the effective Higgs potential (calculated with perturbation theory in the Landau gauge).
The symmetries broken by the Higgs vev are the spontaneously broken symmetries.
\end{defn}

The determination of the Higgs vev by minimizing the Higgs potential is a mean-field approximation\footnote{See~\cite{symmetrybreaking}. Note that Def.~\ref{pt} involves symmetry breaking vevs since perturbation theory can only deal with small perturbations of the Higgs field, which is only guaranteed if the Higgs vev is non-null.
Even considering superselection sectors~\cite{Fredenhagen:2006rv,*Wightman:1990kq,*Wick:1952nb}, we deal with a statistical ensemble of systems each with a non-null Higgs vev corresponding to one superselection sector and we study each system perturbatively.}. Therefore, Def.~\ref{pt} is consistent with Def.~\ref{stat} under the assumption that such mean-field approximation is appropriate, which is often not the case in statistical mechanics and solid state physics\footnote{A simple example where the mean-field approximation predicts spontaneous symmetry breaking in disagreement with the exact solution is the one-dimensional Ising model~\cite{meanfield,criterium,symmetrybreaking}. 
On the other hand, there may be symmetries which we expect to be conserved, but due to yet unknown mechanisms in quantum field theory~\cite{Heissenberg:2015tji}, are in fact spontaneously broken.

The classical problem of minimization of a polynomial ~\cite{Sartori:2003ac,*Sartori:1992ib,*Abud:1983id,stratifiedmorse,*stratifiedmorse2,Wybourne:1980eh,*O'Raifeartaigh:1986vq,*indefinite} is a hard problem where
topology is very useful~\cite{topologypotential,vevacious}. However, to find out which kind of topological transitions can
entail a thermodynamic phase transition is still an open question~\cite{geometrystat}.}. There are many non-exact examples and a rich literature on methods beyond the mean-field approximation~\cite{Grasso:2016gls,Egido:2016bdz}. A common problem in mean-field approximations is the apparent breaking of some symmetries which are in fact conserved by the real system. Restoration of symmetries is crucial to obtain physical states with correct quantum numbers and to account for the quantum fluctuations. E.g. in nuclear physics, the mean-field approximation often clashes with the angular momentum quantum numbers~\cite{Grasso:2016gls,Egido:2016bdz}.

It is well known that for sufficiently strong interactions the perturbative corrections to the mean-field approximation fail to account for the quantum fluctuations~\cite{Osterwalder:1977pc}. This leads many people to extrapolate and assume that for sufficiently weak interactions the perturbative corrections to the mean-field approximation are a good approximation. But the perturbative corrections to the mean-field approximation often fail even at weak coupling:
\begin{itemize}
\item in Quantum Electrodynamics in 2+1 dimensions of space-time (the compact version with a lattice regularization) there is confinement and the absence of a global electromagnetic charge even at weak coupling~\cite{qedpolyakov}, in contradiction with perturbation theory (the same perturbation theory which in 3+1 dimensions of space-time produces precise predictions);
\item the Higgs vev vanishes in the temporal gauge~\cite{Frohlich:1981yi}, while the mean-field approximation predicts a non-null Higgs vev in any gauge;
\item in a simplified lattice model at weak coupling, when the Higgs mass is smaller than the W mass the simulation results are very different from the predictions of perturbation theory \cite{Maas:2013aia,Langguth:1985eu,Evertz:1985fc};
\item it is insufficient to predict the physical spectrum of the $SU(5)$ Grand Unified Theory (Georgi-Glashow)~\cite{Maas:2017xzh} (see also Section~\ref{sec:guts});
\item in a toy model analogous to a Grand Unified Theory at weak coupling there are dramatic differences in the spectrum in the lattice simulation and in the perturbative prediction~\cite{Maas:2016ngo};
\item the mean-field approximation may also fail in superconductors~\cite{Tada:2016lwt}.
\end{itemize}
    
There is a further ingredient to take into account~\cite{Elitzur:1975im}: a spontaneous breaking of local gauge symmetry  without gauge fixing may be impossible in a gauge theory such as the Electroweak theory. The argument is based on the fact that local gauge transformations affect only a small sized system near each space-time point and so there is no infinite size limit as in Def~\ref{stat}\footnote{Under some assumptions on the analiticity  of $\omega_{J,N}$. Note that since the quantum field theory is not well defined mathematically, it is hard to rigorously prove that Elitzur's theorem~\cite{Elitzur:1975im} is valid or that it is not valid~\cite{Damgaard:1985nb,Hsu:1993zc}}.
It can be argued that the Higgs mechanism avoids the presence of Nambu-Goldstone bosons precisely because the local gauge symmetry is not spontaneously broken~\cite{englert,'tHooft:1979bi,*Hooft2007661}. Many non-perturbative studies support this picture~\cite{Osterwalder:1977pc,Fradkin:1978dv,Caudy:2007sf,Seiler:2015rwa}. Moreover, there is a group-theory correspondence between gauge-invariant composite operators and the gauge-dependent elementary fields in the Electroweak theory~\cite{'tHooft:1979bi,Frohlich:1981yi} (also for two-Higgs-doublet models~\cite{Maas:2016qpu}).
 
However, the fact is that the perturbative predictions from the Electroweak theory seem to be a very good approximation to the existing  experimental data in high-energy physics\cite{pdg}, and the (non-perturbative) lattice simulations so far support this picture~\cite{Wurtz:2013ova,Maas:2013aia,Maas:2014pba}. Therefore, it is important to confront these definitions, not only for formal reasons, but also for phenomenological reasons since non-perturbative methods such as lattice simulations~\cite{DeGrand:2015zxa} and the functional renormalization group~\cite{Eichhorn:2015kea} are becoming increasingly relevant in the studies of Electroweak physics and beyond, and are well established in Flavour physics and Quantum Chromodynamics (QCD).

The orbit space approach to the study of invariant functions\cite{Sartori:2003ac,*Sartori:1992ib,*Abud:1983id} provides pure mathematical reasons why explicit and spontaneous symmetry breaking are necessarily related in the context of a (eventually non-renormalizable) potential of arbitrary order. Such kind of relations were noted recently for the CP (charge-parity) symmetry in several multi-Higgs-doublet models and were summarized in the form of a conjecture~\cite{Branco:2015bfb,ivanovseminar} (see also Section~\ref{sec:algebraic}), but for renormalizable potentials which is intriguing.


In this paper we address four problems in the context of extended Higgs sectors:
\begin{itemize}
\item check that the non-perturbative Definition~\ref{stat} of spontaneous symmetry breaking is compatible with the usual assumptions of perturbation theory (Def.~\ref{pt});
\item how to study the spontaneous breakdown of global symmetries without assuming that the local gauge symmetry is spontaneously broken;
\item the relation between explicit and spontaneous symmetry breaking;
\item the relation between the CP symmetry and real representations of the symmetries of the Standard Model.
\end{itemize}

In Sec.~\ref{sec:background},~\ref{sec:real} we state the assumptions we will make throughout the paper,  reviewing background symmetries: these are symmetries that despite they are already explicitly broken, can be still spontaneously broken.
In Sec.~\ref{sec:explicit} we discuss explicit symmetry breaking, so that Def.~\ref{stat} of  spontaneous symmetry breaking applies. Our assumptions and framework are compatible with the usual assumptions of Electroweak symmetry breaking (Def.~\ref{pt}), as we show in Sec.~\ref{sec:spontaneous} and Appendix~\ref{sec:proc}.
In Sec.~\ref{sec:applications} we apply our formalism to extended Higgs sectors and to CP violation. 
We conclude in Sec.~\ref{sec:conclusion}.

\section{Background symmetries of the classical Action}
\label{sec:background}
\subsection{Background symmetries of the functional}
Let $\mathcal{A}$ be a complex algebra of operators, let $G$
be a linearly reductive group\footnote{The class of linearly reductive groups includes all compact groups and the Lorentz group and its double covers that act on the spinors. Such class therefore covers all the groups that are relevant in the Standard Model and in many of its extensions.} of transformations $\mathcal{A}\to \mathcal{A}$, with $G_b$ and $G_f\subset G_b$ normal subgroups of $G$. Note that the local gauge group is always a normal subgroup of the group of symmetries and it is contained in $G_f$, while the quotient group $G/G_f$ is a group of global symmetries.

Consider a $G_f$-symmetric linear functional $\omega:\mathcal{A}\to \mathbb{C}$, by definition all the symmetries conserved by $\omega$ are explicitly conserved by all correlation functions, 
independently of whether the symmetries are spontaneously broken or not. That is, $\omega(A)=\omega(g A)$ for all $A\in\mathcal{A}$ and all $g\in G_f$.

The $G_f$-invariant operators are a representation space of the group $G/G_f$---we have the homomorphism $G\to G/G_{f}$ where  $G_{f}$ is the kernel of the homomorphism.

Consider now the functional $\omega_B$ depending on a  $G_f$-invariant background field\footnote{
We only consider commuting (i.e. non-Grassmann) background fields. A spurion or (non-dynamical) background field enters in the definition of the Lagrangian but it is not a field of the Lagrangian.  When calculating the observables, the background fields are replaced by numerical values. Such numerical values (and the usual fields) are a representation of a group of background symmetries of the classical Action, but there are no Noether's currents associated with such background symmetries if the numerical values are non-trivial. The observables are invariant under the action of the group of the background symmetries. See Ref.~\cite{group2HDM,group2HDM2,*Botella:2012ab,*georgi,Fallbacher:2015rea,Isidori:2012ts} for details and related studies.} $B$, i.e. the expectation value $\omega_{B}(A)$ of the operator $A$ is a (classical) function of the numerical values $B$ and $gB=B$ for all $g\in G_f$. 

In analogy with Def.~\ref{stat}, we say that $G/G_b$ is a background symmetry of $\omega_B$ when
 $\omega_B(A)=\omega_{cB}(c A)$ for all $G_b$-invariant operator $A\in\mathcal{A}$ (by $G_b$-invariant we mean $A$ verifies  $\omega_B(A)=\omega_{gB}(g A)$ for  all $g\in G_b$). We make no assumptions on whether $\omega_B$ is $G_b$-invariant or not.
We do assume that the $G_b$-invariant operators are polynomials  of the fields and of the background
fields\footnote{There is a more general definition: $G/G_{b}$ is a background symmetry of $\omega_B$ when for any $g\in G$ there is some $h\in G_b$ such that $\omega_B(A)=\omega_{g h B}(g h A)$ for all  operator $A$. These two definitions are equivalent when $G_b$ is a linearly reductive Lie group: there is then a finite algebraic basis of $G_b$-invariants parametrizing faithfully the $G_b$-orbit space~\cite{Sartori:2003ac}.

For compact groups we can assume the operators to be smooth functions of the fields and of the background
fields~\cite{compactlieinvariants,SCHWARZ1975,invariantfunctions}.
}.  Note that  the transformation $\omega_B(gA)\to \omega_B(gA)$ affects only the fields (not the background fields) while the transformation  $\omega_B(A)\to \omega_{gB}(A)$ affects only the background field, for any $g\in G$. 

Suppose that $a\in G$ is conserved, then any background transformation $g\in G_b$ modifies the symmetry transformations as $a\to g a g^{-1}$, that is 
$\omega_{gB}(g A)=\omega_B(A)=\omega_{B}(a A)=\omega_{gB}(ga g^{-1} g A)$ (for a $G_b$-invariant operator $A$).

As consequence of the isomorphism theorems~\cite{scott1964group}, the following groups are isomorphic $G/G_{b}\simeq (G/G_{f})/(G_{b}/G_{f})$ and the homomorphism $G\to G/G_{b}$ can be achieved in two steps: first $G\to G/G_{f}$ and then $G/G_{f}\to  (G/G_{f})/(G_{b}/G_{f})$.
This is important since we can build operators invariant under the background group $G_{b}$ using only the operators invariant under the group of symmetries $G_{f}$ that we constructed in a first step.

\subsection{Spontaneous symmetry breaking of background symmetries} The CP symmetry is explicitly broken in the Standard Model, by the phase of the CKM matrix. Promoting such parameter to a background field $B$, we can still spontaneously break the CP background symmetry in extended Higgs sectors. We introduce a background field $J$ which also breaks CP explicitly, and then we have spontaneous breaking of the CP background symmetry $G/G_b$ when\\
$\lim_{J\to 0} (\omega_{J,B}(A)-\omega_{J,c B}(c A))\neq 0$ for some $G_b$-invariant operator $A\in \mathcal{A}$ and $c$ is the CP transformation, i.e. the generator of the CP group $G/G_b$.

We can consider analogous situations with other compact groups $G$. Using a second background field $J_2$ we could even study the spontaneous symmetry breaking of a symmetry that is already spontaneously broken via $J$.
Therefore, the use of background fields allow us to address a wide class of problems.

For instance, the soft symmetry breaking terms are very useful for phenomenological applications~\cite{Branco:2011iw}. These are quadratic terms of the Higgs potential,
the corresponding parameters can be promoted to background fields, such that the symmetry which is softly broken is a background symmetry. We can therefore study spontaneous symmetry breaking in the context of softly broken symmetries. 

\subsection{Classical Action in Quantum Field Theory} The introduction of Grassmann (anti-commuting) variables to describe fermions in the classical Action, allows us to describe in principle any quantum non-abelian gauge field theories by its classical Action. Such classical Action is well defined mathematically~\cite{Berezin:1976eg,*Casalbuoni:1975bj,*classicalspin}.
While we could consider fermionic background fields, all the parameters of the classical Action (e.g. the Yukawa couplings) as well as the results of the correlation functions are commuting numbers and so it  suffices to consider commuting background fields, as we will do here. 

However, the symmetries conserved by the classical Action may not be conserved by the path integral measure and so by the vacuum functional (i.e. by the full quantum system): we would have then a quantum anomaly~\cite{Geng:1988pr,*fujikawa2004path}. 
Explicit symmetry breaking is different from quantum anomalies: in explicit symmetry breaking the classical action contains explicit symmetry breaking terms, such that when those terms are null both the classical action and the vacuum conserve the symmetry.

Since spontaneous symmetry breaking can be defined as a particular case of explicit symmetry breaking, it is also different from quantum anomalies.  Background symmetries are a particular case of explicit symmetry breaking, as well.

We will study here the action of the group $G$ on the classical Action and assume that the Action, the field content of the theory and the group $G$ are chosen such that the path integral measure is $G$-invariant: i.e. there are no quantum anomalies\footnote{Such assumption is valid for the gauge symmetries of the Standard Model~\cite{Geng:1988pr,*fujikawa2004path} and many extensions, but not for the baryonic number in the Standard Model~\cite{Dine:2003ax}.}.
In any case, the determination of the symmetries of the classical Action is a first step towards the determination of the symmetries of the vacuum, therefore our results are also useful in the presence of quantum anomalies\footnote{The presence of quantum anomalies implies that the study of the symmetries of the vacuum must address also the path integral measure which still lacks a non-perturbative mathematical definition~\cite{prize}, and so can only be done using further assumptions appropriate for each particular quantum theory in separate.}.

Note that since the probability is the modulus squared of the probability amplitude, there may be discrete symmetries of the probability which are not symmetries of the probability amplitude. The time-reversal is one example. We will not consider such symmetries here.

\subsection{Effective Action and ultra-violet incomplete theories}

The renormalization group affects the parameters of the classical Action and thus we need to pay attention on how we define the background fields. We follow the convention used in Minimial Flavor Violation~\cite{Isidori:2012ts}: the background fields are not modified by the renormalization group, only the parameters of the Action are modified. 
The background symmetries are conserved by the renormalization group. However, we need to consider the parameters to be the most general function of the background fields (constrained by the background symmetry of the Action). In practice this means that the parameters will be a simple polynomial of the background fields only at some fixed energy scale at our choice. Once we do the renormalization group running, the parameters will no longer be simple polynomials of the background fields. This justifies that the background fields and the parameters are different, and they should be given different names (so spurion instead of background field is ok, but reparametrization symmetries instead of background symmetries may be misleading).

We also assume that the classical Action is a real polynomial of arbitrary order on the fields.
From the point of view of classical field theory there is no reason to limit the order of the polynomial.
When taking into account the quantum effects, then we are working in the framework of an effective field theory, without making assumptions about the ultra-violet completion of the theory\footnote{The claim that a quantum field theory is ultra-violet complete (i.e. renormalizable) just because the classical Action is a fourth order polynomial has some problems: 
without an ultra-violet cutoff (or some alternative regularization), quantum non-abelian gauge field theories still lack a non-perturbative mathematical definition~\cite{prize}; the perturbative approach to the Standard Model is based on the $\lambda \phi^4$ quantum theory (mexican hat potential), but in the $\lambda \phi^4$ quantum theory the (non-perturbatively) renormalized coupling $\lambda$ seems to be necessarily null (trivial)~\cite{nonperturbativefoundations,triviality2}, the triviality can be avoided with an ultra-violet cutoff and an upper bound for the Higgs mass~\cite{triviality}; the advantage of a logically consistent 
ultra-violet complete theory over one incomplete theory would be to explain the meta-stability of the Standard Model in face of the present experimental data~\cite{Eichhorn:2015kea,Degrassi:2012ry} and quantum gravity (at the Planck scale), no such theory is yet known.
 
Note that extensions to the Standard Model often  have so much degrees of freedom that either we set the non-renormalizable interactions to zero or not, they are anyway phenomenologically relevant. E.g. the renormalizable two-Higgs-doublet model has enough degrees of freedom to emulate a standard Higgs sector with free couplings at the LHC (ignoring the non-LHC constraints)~\cite{smfree}; while the non-renormalizable two-Higgs-doublet model is also relevant~\cite{Crivellin:2016ihg}.

We can have in principle constant fields without space-time dependence, for which renormalizability imposes no restriction on the order of the polynomial. Or a time-independent problem, where the restriction is different from fourth order.
}. Surely, we require the classical Action to be such that the quantum theory is predictive enough and valid up to an interesting energy scale: implying that higher order interactions should be fewer and much smaller~\cite{Sher:1988mj} (at the electroweak energy escale), but not necessarily a fourth order (or any other limit on the order of the) polynomial.
For instance, the two-Higgs-doublet model can be formulated as an effective field theory~\cite{Crivellin:2016ihg}.

The presence of a cutoff scale is essential to allow for taking the limit of vanishing explicit symmetry breaking: the divergences in perturbation theory may be large, but they are always finite due to the cutoff. Therefore the limit of vanishing explicit symmetry breaking is well defined even in the presence of divergences.

Note that in many practical cases (for instance~\cite{Maas:2016qpu}), we can define spontaneous symmetry breaking as a particular case of explicit symmetry breaking within a renormalizable potential.
In fact, if the gauge symmetry is $SU(2)_L\times U(1)_{em}$ (as in multi-Higgs-doublet models) we can always do it, at the cost of eventually breaking explicitly more global symmetries than we would in a non-renormalizable potential (for an arbitrary gauge group that is not the case, since the invariant tensors can involve polynomials of order$>4$). But the presence of a cutoff is anyway essential as explained previously, even in a renormalizable potential.

Note that we could consider other definitions of spontaneous symmetry breaking, where the cutoff is not essential. But the other definitions also have problems to be solved. For instance, boundary conditions can be implemented in principle, but we need to implement them at the (non-perturbative) quantum level.

\section{The fields are real representations of the group of symmetries}
\label{sec:real}
 
In the canonical quantization of free fields, the charged scalar field is built from two real scalar fields~\cite{aqft,Aranda:2016qmp}.
We have for the real (i.e. self-adjoint) fields:
\begin{align*}
\varphi_{r,i}(\vec{x},t)=\int \frac{d^3\vec{p}}{(2\pi)^3 2 E_p} a_{r,i}(\vec{p})e^{-ip\cdot x} + a^\dagger_{r,i} (\vec{p})e^{ip\cdot x}
\end{align*}

Where the subscript $r,i$ stands for real or imaginary parts.
The charged scalar field is a 2-dimensional real representation of the $U(1)_{em}$ gauge group,
it is composed of 2 real scalar fields, i.e. $(\phi_r,\phi_i)$.

But since the algebra of fields is complex\footnote{In a more elaborate treatment the algebra of operators can be chosen as a real algebra~\cite{realoperatoralgebras,realpoincare}. However, for the purposes of this work it suffices to consider a complex algebra of operators.}, this allows to rewrite the charged scalar field as:
\begin{align*}
 \varphi=\frac{1}{\sqrt{2}}(\varphi_{r}+i \varphi_{i})
\end{align*}

Which is the same as the usual form:
\begin{align*}
\varphi(\vec{x},t)=\int \frac{d^3\vec{p}}{(2\pi)^3 2 E_p} a(\vec{p})e^{-ip\cdot x} + b^\dagger (\vec{p})e^{ip\cdot x}
\end{align*}

Where $a=\frac{1}{\sqrt{2}}(a_{r}+i a_{i})$ and $b^\dagger=\frac{1}{\sqrt{2}}(a^\dagger_r+i a^\dagger_{i})$. This notation seems to imply that the charged scalar field is instead a complex representation of the $U(1)_{em}$ gauge group.
But that is not true because charge conjugation $C$ is an unitary transformation: it commutes with the imaginary number and acts
as $C\varphi C^\dagger=e^{i\theta}\varphi^*$, where $\theta$ is an arbitrary phase\footnote{This is the most general $C$ transformation for an irreducible representation of $U(1)_{em}$.
  The most general $C$ transformation for reducible representations of $U(1)_{em}$ will be discussed in Section~\ref{sec:algebraic}}.

Thus, the charged scalar field conserves an anti-linear condition\footnote{Such condition is related with the time-reversal transformation, but we do not need to discuss time-reversal here.}:
$(C \varphi C^\dagger)^*= e^{-i\theta}\varphi$. Therefore, linear combinations with complex coefficients of the field operators do not conserve the anti-linear condition: e.g. $\varphi(\vec{x},t)+i\varphi(\vec{y},t)$ does not conserve the anti-linear condition (where $\vec{x},\vec{y}$ are arbitrary positions). This implies that the charged scalar field is a real representation of the group of symmetries.

Another way to see this is that we can write the real form
$2(\varphi_r,\varphi_i)=(\varphi+C\varphi C^\dagger, -i\varphi +C \varphi C^\dagger)$ using linear transformations of the complex form $\varphi$.
Thus, whether we write the scalar field as $(\varphi_r,\varphi_i)$ or as $\frac{1}{\sqrt{2}}(\varphi_{r}+i \varphi_{i})$, we are still dealing with a real representation of the group of symmetries.
This would not happen if the charge conjugation would be an anti-unitary transformation, i.e. involving a complex conjugation and thus anti-commuting with the imaginary number.
In that case, $\varphi$ would be a complex representation.

The extension of this argument to other (not scalar) bosons is straightforward. For fermions, a similar thing happens. We can see that $a=\frac{1}{\sqrt{2}}(a_{r}+i a_{i})$ and
$b^\dagger= \frac{1}{\sqrt{2}}(a^\dagger_r+i a^\dagger_{i})$ verify the anti-commutation relations
$\{a,a^\dagger\}=1$ and $\{b,b^\dagger\}=1$ and $\{a,b^\dagger\}=\{a,b\}=0$, assuming that $\{a_r,a^\dagger_r\}=1$ and $\{a_i,a^\dagger_i\}=1$  and $\{a_r,a^\dagger_i\}=\{a_r,b_i\}=0$.
Thus a (complex) Dirac spinor field is built from two (real) Majorana spinor fields.
Since, the charge conjugation is a linear operation, also the fermions are real representations of the group of symmetries.

In classical field theory, the fields can be complex representations\footnote{In the case that the fields are complex representations, the charge conjugation is anti-linear as in reference~\cite{realizable}. But in Quantum Field Theory the charge conjugation is anti-linear as was recognized in Ref.~\cite{Aranda:2016qmp} which shares a common author with the reference~\cite{realizable}.} or real representations at will. It may be tempting to consider complex representations, since Holomorphic functions (i.e. $\frac{\partial f(z,z^*)}{\partial z^*}=0$ where $z^*$ is the complex conjugate of the complex variable $z$)
are the central objects of study in complex analysis which has many applications.
However, the classical Action is not an holomorphic functional:
$\frac{\partial S(\phi,\phi^*)}{\partial \phi^*_x}\neq 0$ for any complex field $\phi$.

Therefore, there is no advantage a-priori in the fields and background fields being complex vector spaces, even in classical field theory:
$S(\phi,\phi^*)=S(Re(\phi),Im(\phi))$. 
For instance, the orbit space methods~\cite{Sartori:2003ac} are valid for real or complex vector spaces; the Wigner theorem is valid for unitary/anti-unitary representations on complex vector spaces as well as for real orthogonal (i.e. unitary) representations on real vector spaces~\cite{Ratz1996,*wignertheorem}.

On the one hand, complex irreducible representations of the group $G\times H$ are a direct product of complex irreducible representations of $G$ and of $H$, which is not the case for real irreducible representations. On the other hand, with complex representations we cannot define all linear operators available in Quantum Field Theory, some of these operators have important experimental consequences, for instance the approximate custodial symmetry can only be defined when the Higgs field is a real representation of $SU(2)_L$~\cite{pdg,Pich:2015lkh,Pilaftsis:2016erj} or the Majorana condition on fermions cannot be defined~\cite{PhysRevLett.116.257003}.

Note that the standard notation is intuitive and practical, using complex fields to represent charged fields. But we have shown that despite intuitive, such notation does not imply that the fields are complex representations. On the contrary, the standard approach is to treat the fields as real representations, which is expected since the classical Action really is a classical functional in the sense of Classical Field Theory and there is no notice  that complex fields are indispensable in Classical Field Theory~\cite{Berezin:1976eg,*Casalbuoni:1975bj,*classicalspin}.

\section{Explicit symmetry breaking}
\label{sec:explicit}

Without knowing much about our system, we can classify the explicit symmetry breaking by a background field $J$.
We will call $J$ the source field, to distinguish it from the remaining background fields. 

As in the previous section, $G$ is a linearly reductive group, with $G_b$ and $G_f\subset G_b$ normal subgroups of $G$.

The background symmetry $G$ is by definition always explicitly conserved in the absence of the source field.
Therefore, when we refer to the explicit breaking of a background symmetry we mean in the presence of the source field.

There are then 2 different possibilities for the source field:

\paragraph{1) $J$ breaks the background symmetry $G/G_b$} 

The breaking term may be non-renormalizable. For instance, for one complex scalar field, $G_f=Z_6$ and $G=U(1)$ we need a non-renormalizable potential to break $U(1)$ while conserving $Z_6$.

\paragraph{2) $J$ conserves the background symmetry $G/G_b$}

This means that $G J=G_b J$.
Then, for every $g\in G$ there is $h\in G_b$ such that $J= g h J$.




If $G/G_b$ is not just a background symmetry but it is also a true symmetry of the Action, there are two further possibilities.



\paragraph{2.A) $J$  conserves the symmetry $G/G_b$} 
Then we can find $h$ such that $g h J = J$ and $gh$ is conserved by the Action.

\paragraph{2.B)  $J$ breaks the symmetry $G/G_b$}

Note that we can always find $h,k\in G_b$ such that $g k$ is conserved by the Action and
$h g J = J$. Thus we have that $g k J \neq J$ holds necessarily.

\section{Spontaneous symmetry breaking due to the Higgs potential}
\label{sec:spontaneous}

The only difference with respect to the usual perturbative expansion is that we only evaluate vevs of gauge-invariant operators so we do not assume that the gauge symmetry is spontaneously broken.

The Higgs potential has a symmetry $G/G_b$ and $G_f$. The subgroups $G_b$ and $G_f\subset G_b$ are normal subgroups of $G$ which is a compact group. We also assume that $G/G_f$ is a group of global transformations\footnote{In the case of the Standard Model, since the $U(1)_{Y}$ gauge symmetry is abelian and there are no Gribov-Singer ambiguities for abelian gauge fixing (unlike for a non-abelian gauge symmetry such as $SU(2)_L$), we can unambiguously fix the local gauge with a gauge-fixing local term and deal only with the $U(1)_Y$ global symmetry.}.

In analogy with Def.~\ref{pt}, to study the Higgs potential and in particular the global symmetries which are spontaneously broken or not, we assume that: 
\begin{itemize}
\item we fix the local gauge and calculate the effective potential, using
the standard methods: Landau gauge-fixing, perturbative expansion in the number of loops and renormalization group running~\cite{Sher:1988mj,Lee:1974fj,Coleman:1973jx,Weinberg:1973am});
\item We then assume that the vevs of operators are given by the usual perturbative expansion, which is an expansion in
1) the Higgs field around one point  $\frac{v}{\sqrt{2}} \phi_0$ (constant in space-time and $\phi_0^\dagger\phi_0=1$) for which the effective potential has an absolute minimum; and 2) the couplings of the interactions\footnote{The usual perturbative expansion is an expansion in the couplings with the mass of the $W$ boson  kept finite ($M_W=g v/2$) therefore it is also an expansion for large Higgs vev.  

The vevs of the local gauge invariant operators are the physical observables if the local gauge symmetry is not spontaneously broken, as it seems to be the case~\cite{englert,Frohlich:1981yi,'tHooft:1979bi}.
In the context of the perturbative formulation of Electroweak theory, there are already studies of the (multi-)Higgs potential based on $SU(2)_L$-invariant bilinears of the Higgs field~\cite{accidental2,accidental,Pilaftsis:2016erj,Nishi:2011gc}.

The local gauge fixing is perturbative with a local term and in a suitable gauge~\cite{Frohlich:1981yi,Maas:2012ct} (such as the usual gauges used in perturbation theory), we assume that the (non-perturbative) Gribov-Singer ambiguities do not affect our results. The reference point is constant in space-time with respect to the chosen local gauge.}; we consider the set $\{V,\phi_0\}$ formed by the Higgs potential and the reference point minimizing the Higgs potential
\item whenever there are two or more $G_b$-orbits of $\{V,\phi_0\}$ (i.e.
  $\{\{h V, h \phi_0\}: h\in G_b\}$), then there is spontaneous symmetry breaking of the background symmetry $G/G_b$ if these $G_b$-orbits are related by $G$.
\item we assume that the non-perturbative vevs of operators conserve the same symmetries as the vevs calculated in perturbation theory, i.e. the symmetries conserved by the classical Action and the reference point~\cite{Sher:1988mj}.
\end{itemize}


These are non-trivial assumptions, requiring that the mean-field approximation describes spontaneous symmetry breaking correctly. In the Standard Model, the $SU(2)_L$-gauge orbit minimizing the Higgs potential is unique and therefore there is no experimental evidence in the context of Electroweak physics, that these assumptions relating spontaneous symmetry breaking with non-unique $SU(2)_L$-gauge orbits are valid.
Thus more studies, simulations and experimental data are required to support these assumptions~\cite{Lewis:2010ps,Maas:2016qpu}. Even if these assumptions are valid, it is still not an easy task to determine if some symmetry is spontaneously broken or not by the radiative corrections, if it is conserved at tree-level~\cite{Georgi:1974au,Carena:2000yi,Branco:2000dq}. Note however that the radiative corrections do conserve the background symmetries explicitly, in the absence of quantum anomalies in the measure of the path integral.

Under these assumptions, we have to solve a classical (but still non-perturbative) problem of minimization of a polynomial invariant under a group of symmetries~\cite{stratifiedmorse,*stratifiedmorse2,Wybourne:1980eh,*O'Raifeartaigh:1986vq,*indefinite}.

If there is spontaneous symmetry breaking of $G/G_b$
under these assumptions, then we modify the tree-level Higgs 
potential: we add an infinitesimal polynomial conserving the symmetry $G_f$
which has a $G_b$ background symmetry  (defined by a source field $J$) and such that the $G_b$-orbit of the set $\{W,\phi_0\}$ (formed by the effective modified Higgs potential $W$ and a reference point minimizing $W$) is not related by $G$ to a set defined by another minimum of $W$~\footnote{Note that the reference point $\phi_0$ can still be not be the unique minimum, but this arbitrary choice of reference point is not related to $G/G_b$. 

  Also, it does not suffice to explicitly break the symmetry in the classical Action without affecting the minimum of the modified effective potential, otherwise it would be possible to have vevs breaking the symmetry $G$ even when the explicit symmetry breaking term is exactly null which would be inconsistent with Def.~\ref{stat}).}. Then the modified potential explicitly breaks $G/G_b$. If perturbation theory is correct there are finite vevs breaking $G/G_b$ in the limit that the modified potential converges to the Higgs potential and so there is spontaneous symmetry breaking. However, our procedure is still valid in case perturbation theory fails, since what we determined were the symmetries of the (infinitesimally) modified tree-level potential.

In Appendix~\ref{sec:proc} we show that it is always possible to modify the potential by an infinitesimal term 
such that its minimum is the one we want. If explicit breakdown of the symmetry $G/G_b$ is not possible then for all $\phi$ and $g\in G$ there is always $h\in G_b$ such that $gh \phi=\phi$ and the potential conserves $gh$; then spontaneous symmetry breaking due to the Higgs field is not possible.

Note that the $G_f$-invariants are a faithful representation of the group $G/G_f$~\cite{Sartori:2003ac,*Sartori:1992ib,*Abud:1983id} and thus allow us to study if the global symmetries
$G/G_f$ are spontaneously broken, without making assumptions on whether the gauge symmetry is conserved or spontaneously broken.
Therefore, these assumptions are compatible with further assumptions on gauge symmetry breaking, and they are suitable for studies looking for evidence of the spontaneous breaking of the gauge symmetry---e.g. comparing perturbative predictions from vevs of gauge-invariant/dependent operators with experimental results and with non-perturbative studies.

The electromagnetic symmetry $U(1)_{em}$  is the representation of the $U(1)_Y$ gauge symmetry in the $SU(2)_L$-gauge-invariant operators and we can treat it as a global symmetry after $U(1)_Y$ local gauge fixing.
Therefore, under our assumptions the  $U(1)_{em}$ symmetry can also be spontaneously broken like all other global symmetries if there are two $SU(2)_l$-gauge-orbits minimizing the Higgs potential related by a $U(1)_{em}$ transformation\footnote{We are not dependent on these assumptions to determine what would happen if the $SU(2)_L$-gauge orbit minimizing the Higgs potential breaks the $U(1)_{em}$ generator:  the photon would become massive due to the abelian Higgs mechanism---there are theoretical arguments~\cite{DePalma:2013kua} and also experimental evidence from superconductivity where the abelian Higgs mechanism also happens. It would not depend on the $U(1)_Y$ gauge-fixing and would not imply spontaneous breaking of the local gauge $U(1)_{em}$~\cite{symmetrybreaking,Tada:2016nve}. The $U(1)_Y$ gauge-fixing merely allows us to simplify the study by treating the $U(1)_Y$ symmetry and the remaining global symmetries in the same consistent way, which is particularly useful to interpret the results of non-perturbative lattice studies where $U(1)_Y$ is not a local gauge symmetry (reducing computation time)~\cite{Wurtz:2013ova,Maas:2013aia,Maas:2014pba}.}.

\section{Applications}
\label{sec:applications}

\subsection{The problem with Grand Unified Theories}
\label{sec:guts}

Recently, it was pointed out that the predictions of perturbation theory in the Georgi-Glashow $SU(5)$ Grand Unified Theory are probably inconsistent~\cite{Maas:2017xzh}.

The construction of a non-abelian charge requires dressing the gauge-dependent elementary fields, forming local gauge invariant operators. If the construction of non-abelian charges would be possible, then such construction would be a better alternative to the local gauge-fixing condition (used in perturbation theory), which is globally ill-defined due to the Gribov ambiguity.
However, the only known ways of dressing the fields, require the existence of a local gauge-fixing condition which is globally well defined~\cite{Lavelle:1995ty}. Thus no such alternative to local gauge-fixing is known and so assuming the existence of a non-abelian charge is speculative, even at weak coupling~\cite{Maas:2016ngo}.

The above discussion implies that the global electromagnetic charge is difficult to reproduce in a Grand Unified Theory, while the Standard Model allows to reproduce the electromagnetic charge since there is no Gribov ambiguity in abelian gauge-fixing .
There is thus a structural difference between the perturbation theory and the gauge-invariant formulation, as was already noted in reference~\cite{Maas:2017xzh}.

In the following we discuss what are the requirements for the predictions of perturbation theory in the Georgi-Glashow $SU(5)$ Grand Unified Theory to be consistent.
The conclusion will be that these requirements are significant and untested, in other words the fact that perturbation theory works in the Standard Model does not suggest in any way that it works in the $SU(5)$ Grand Unified Theory.

The invariant tensors of the $SU(5)$ group are algebraic combinations of the Kronecker delta and the Levi-Cita tensor.

We consider two Higgs fields ($\Phi,\Sigma$) in an irreducible (fundamental,adjoint) representation of $SU(5)$.

After gauge-fixing, we consider that the vevs of the two Higgs fields are \\
$<\Phi_j>=\delta_{j5},\quad <\Sigma>=diag(2,2,2,-3,-3)$.

In that case, we can form operators whose vacuum expectation values are projections for the different representations of $SU(3)$  and $SU(2)$ or $U(1)_{em}$, following the procedure of reference~\cite{Frohlich:1981yi}.

So, contracting with the Higgs field $\Phi$ will create a trivial representation of  $SU(3)$  and $U(1)_Y$.

Applying the projection $P_0=\Phi \Phi^{\dagger}$ will create a trivial representation of $U(1)_{em}$ and  $SU(3)$.

Applying the projection $P_+=\frac{2-\Sigma}{5}-P_0$ will create a charged state of $U(1)_{em}$ and trivial representation of  $SU(3)$.

Applying the projection
$P_3=\frac{3+\Sigma}{5}$ will create a fundamental representation of $SU(3)$ and trivial representation of $SU(2)$.

Applying the projection
$P_2=1-P_3=\frac{2-\Sigma}{5}$ will create a fundamental representation of $SU(2)$ and trivial representation of $SU(3)$.

Note that $P_++P_0=P_2$.

Then $Re(P_2D_\mu P_3)$ and $Im(P_2D_\mu P_3)$ account for 12 massive vector bosons.
These can be divided into 6 bosons $P_0 D_\mu P_3$, 
and 6 bosons $P_+ D_\mu P_3$.

Also $(P_3\Sigma P_3-\frac{1}{3}\tr(P_3\Sigma P_3) 1)$, $(P_2\Sigma P_2-\frac{1}{2}\tr(P_2\Sigma P_2) 1)$, $(\frac{2}{3}\tr(P_3\Sigma P_3)-\frac{3}{2}\tr(P_2\Sigma P_2))$ ,  account for $8+3+1=12$ Higgs bosons.
These 3 bosons $(P_2\Sigma P_2-\frac{1}{2}\tr(P_2\Sigma P_2) 1)$ can be divided into 2 bosons $(P_+\Sigma P_0)$ plus 1 boson $(P_+\Sigma P_+-P_0 \Sigma P_0)$.

The 12 Higgs bosons plus the 12 goldstones (present in the massive vector bosons) correspond to the 24 degrees of freedom of $\Sigma$.

Then $P_+ D_\mu \Phi$ and $\Phi^\dagger D_\mu \Phi$ account for 3 massive vector bosons ($W^+$ and $W^0$).

Also $P_3\Phi$, $\Phi^\dagger\Phi$ account for $6+1=7$ Higgs bosons.

The 7 Higgs bosons plus the 3 goldstones (present in the massive vector bosons) correspond to the 10 degrees of freedom of $\Phi$.

The fermions can be easily obtained using the projections $P_3$, $P_+$ and $P_0$ on the $SU(5)$ fermionic fields.
It remains to solve how to obtain local gauge-invariant states of $SU(3)\times U(1)_{em}$.

Now we need an
``effective'' $U(1)_{em}$ global charge. Following the suggestion of the reference~\cite{Frohlich:1981yi}, such global charge is created effectively, from the interaction of (for instance) an electron located at an infinite distance, multiplied by the parallel transport needed to achieve gauge invariance.

Such state is $P_4\Psi$ where $\Psi$ is an $SU(5)$ fundamental representation containing the electron and the projection $P_4$ will select the electron.

With such non-trivial assumption, then perturbation theory works since we can generate all invariant tensors of  $SU(3)$  and $U(1)_{em}$ from invariant tensors of $SU(5)$.

To make a $U(1)_{em}$ charged state invariant of $SU(5)$ we contract it with $P_4\Psi$.
We have the $SU(3)$ invariant tensors Levi-Civita  $\epsilon_{abcde}(P_4\Psi)^d \Phi^e$ and Kronecker delta $P_3$.

We conclude that contractions with an (effective) electron located at an infinite distance are needed to make perturbation theory work in the $SU(5)$ GUT, and then such contractions can not affect the spectrum of the theory. This a non-trivial and untested requirement beyond the Higgs mechanism of the Standard Model.

\subsection{The unbroken CP symmetry in a toy model}
\label{sec:algebraic}

Recently, in the context of a toy model~\cite{Ratz:2016scn} the CP (charge-parity) symmetry was defined as involving a complex conjugation in a mathematically strict way. Such definition led to inconsistencies in the toy model of reference~\cite{Ratz:2016scn} (even accepting that the gauge symmetry is spontaneously broken) which were overlooked.  As was mentioned previously, since the CP transformation is unitary, the Higgs field is a real representation of the group of symmetries (including CP). Thus, CP can only be related to complex conjugation as a notation convenience, but not in a mathematically strict way.

The toy model of reference~\cite{Ratz:2016scn} after spontaneous symmetry breaking, features a ``complex'' scalar which is left invariant by the CP transformation. This is unusual, but not inconsistent since 
the ``complex'' scalar is in fact a 2-dimensional real representation and CP needs not to complex conjugate it.

The toy model  has the merit of being a clear example that indeed CP should not be defined with respect to complex conjugation. In such toy model, CP is conserved by the Lagrangian and the Higgs vev. The CP-odd invariants (by definition) therefore can only indicate that CP is conserved. Nevertheless, in reference~\cite{Ratz:2016scn} it is claimed that there are  CP-odd invariants which break CP; but the ``invariants'' which were calculated are not gauge-invariant and thus are not  CP-odd invariants (an $SU(3)$ transformation could transform some $W$ into a $Z$ but the ``invariants''  depend on $W$s and not on the $Z$, where the $W$s and $Z$ are defined in reference~\cite{Ratz:2016scn}).

Note that there is a certain freedom in how to define a ``physical CP transformation'' in a toy model. But in any case the CP group (i.e. $G/G_b$ group) must be a group of global transformations and so the CP-odd  invariants (i.e. invariants under the $G_b$ background symmetry and treating the Higgs vev as a background field) must be invariant under the background gauge group. The main claim of reference~\cite{Ratz:2016scn}, namely that there is a toy model exhibiting physical CP-violation while the CP symmetry is conserved by the Lagrangian and the vacuum is false, because it is based on an explicitly false statement about the CP-odd invariants.

\subsection{Accidental symmetries}
\label{sec:pseudo}

Recently, a conjecture was made relating explicit and spontaneous CP violation, but in a renormalizable Higgs potential~\cite{Branco:2015bfb,ivanovseminar}. 

Such conjecture builds on different assumptions than ours. For instance, in the case mentioned in reference~\cite{Branco:2015bfb}, the Higgs potential is invariant under the group $A_4$ and it is not possible to break CP explicitly in a renormalizable $A_4$-symmetric potential. However, because the background group orbit does not absorb all the CP-violating phases (i.e. not all the $G_b$-invariants are $CP$-invariants, where $G_b$ contains $A_4$ as a normal subgroup and $G/G_b=Z_2$ is the CP group), the potential $V(\phi)=-\phi^\dagger\phi+(\phi^\dagger\phi)^2$ allows us to choose an arbitrary minimum which can break the background CP symmetry. Therefore, there are counter-examples to the conjecture that there is a fundamental relation between explicit and spontaneous CP violation in a renormalizable Higgs potential.
The question then is, what is the relevance of these counter-examples?

The authors consider the cases with pseudo-golstone bosons as ``non-physical'' (which is the case of the counter-example discussed above, since the symmetry of the Higgs potential can be explicitly broken in the remaining of the Lagrangian).

The pseudo-Goldsone bosons arise by the breaking of continuous symmetries of the potential which are not symmetries of the Lagrangian. Such accidental symmetries can only be properly identified if the Higgs field is considered to be a real representation of the group of symmetries~\cite{accidental,accidental2,Pilaftsis:2016erj}.  To apply the Georgi-Pais theorem, which excludes radiative CP violation we need to avoid pseudo-Goldsone bosons~\cite{Georgi:1974au}. One important example of an accidental symmetry in the Standard Model is the custodial symmetry, which can only be  identified if the Higgs field is considered to be a real representation of the group of symmetries.

Since there are models featuring pseudo-Goldstone bosons in the literature
\footnote{The case for the two-Higgs-doublet model with a $Z_2$ flavor symmetry softly broken~\cite{Chang:1994mz} or pseudo-goldstone bosons playing the role of axions~\cite{axion}. In several references (e.g. \cite{Branco:2011iw, realizable, Ivanov:2007de} of the same authors of the conjecture~\cite{Branco:2015bfb}, the emphasis is always the symmetries of the full Lagrangian and not the accidental symmetries of the Higgs potential, and so in these references the pseudo-golstone bosons do not seem to be considered as ``non-physical''.} whose authors consider them to be ``physical'', labeling  pseudo-golstone bosons as ``non-physical'' is not at all obvious and we do not do it here.

Thus, under our assumptions, there is indeed a fundamental relation between explicit and spontaneous CP violation but only in a non-renormalizable Higgs potential. Note however that the examples presented in the conjecture~\cite{Branco:2015bfb,ivanovseminar}  were most helpful for this work.

\subsection{Generalized basis for generalized CP}
\label{sec:algebraic}

As we have seen, the Higgs field is a real representation of the group of symmetries.
So, what are we really searching for when we are searching for $CP$-violating phases?

Consider a 3-Higgs-doublet model~\cite{Ivanov:2015mwl,Aranda:2016qmp}. Each Higgs doublet has 4 components. Under $Z_2$ given by the complex conjugation we have for the imaginary neutral component $\phi_i\to -\phi_i$ and for the imaginary charged component $\varphi_i\to -\varphi_i$. We can instead introduce a real background field $\epsilon$ 
with numeric value $1$ or $-1$, which transforms under the $Z_2$ as $\epsilon\to -\epsilon$. Then we can redefine the fields as $\varphi=R^\dagger\phi$ with $R$ such that for each Higgs doublet $\phi_i'= \epsilon\phi_i$ and  $\varphi_i'= \epsilon\varphi_i$.  
Then $\phi_i'$ and $\varphi_i'$ is $Z_2$-invariant but $\epsilon$ may appear in the parameters of the potential, since $\epsilon$ is real.

Then, a potential where $\epsilon$ is not present is CP-conserving. This may seem trivial, but the CP-transformation needs not to be the complex conjugation. It can be given by $Uc$ where $c$ is the CP-conjugation and $U$ is an element of the group of background symmetries~\cite{brancocp,Chen:2014tpa}. The reason is the fact that group extensions are not unique~\cite{extensions,pin,realizable}.

Suppose that we start by imposing a family group $H_f$. Then the CP transformation $Uc$ conserves $H_f$ and so does $(Uc)^2$.
If the CP transformation is conserved then $(Uc)^2=UU^*\in G_f$. So, $G_f$ contains $H_f$ as a normal subgroup. The CP transformation $Uc$ conserves $G_f$ since $Uc$ commutes with $UU^*$.
 
After we identify all the symmetries $G_f$ of the system which commute with $U(1)_Y$, to check if $CP$ is conserved or not we need to check if it breaks the $G_f$-invariants. But the CP transformation always acts on the $G_f$-invariants as a $Z_2$ transformation since $(Uc)^2=UU^*\in G_f$.

So, we start with the same background field $R$, it is invariant under $G_g\times U(1)_{em}\times SU(2)_L$.
However, under a generalized CP transformation we get the background field $R\to UR^*$. Note that $R R^\dagger=1=R^\dagger R$ is still left invariant by the generalized CP transformation, so we can insert $R^\dagger R$ wherever it is necessary.
Also $UR^* G_f\phi=G_fUR^*\phi$ still conserves the true symmetries $G_f$, despite that it changes the remaining background symmetries. 

Then, we change the basis of the potential using $R$, we call it the $CP$-basis. 
The parameters of the Higgs potential in the $CP$-basis (e.g. $\mathcal{Y}=R^\dagger Y R$ and  $\mathcal{Z}=(R^\dagger\otimes R^\dagger) Z (R\otimes R)$) are by construction $G_f$-invariants. In particular, $(cU)^2=U^*U\in G_f$ and so the $CP$ transformation acts on the parameters of the Higgs potential as a $Z_2$-transformation.

Therefore in such basis, we still need to look for terms depending on $\epsilon$. If there are none, then $CP$ is conserved and in that sense we would have a ``real'' basis.
Note however that it may not be practical to find such basis because $R$ can be in principle any unitary matrix. But such ``real'' basis always exists.

In the case of an order-$4$ CP transformation $Uc$ with $U=\left[\begin{smallmatrix}
1 & 0 & 0\\
0 & 0 & -i\\
0 & i & 0
\end{smallmatrix}\right]$~\cite{Ivanov:2015mwl,Aranda:2016qmp}. 
We have $R=\left[\begin{smallmatrix}
1 & 0 & 0\\
0 & a & -b\\
0 & b & a
\end{smallmatrix}\right]$ and $a=1\to 0\to -1 \to 0$ and $b=0\to i\to 0 \to -i$, where the arrows indicate the action of the CP.

Then for the $UU^*$-invariants, 
we have\\
 $\phi_2^\dagger \phi_2=a^2\varphi_2^\dagger \varphi_2+bb^*\varphi_3^\dagger \varphi_3$,\\
  $\phi_3^\dagger \phi_3=b b^*\varphi_2^\dagger \varphi_2+a^2\varphi_3^\dagger \varphi_3$,\\
$\phi_3^\dagger \phi_2=a^2\varphi_3^\dagger \varphi_2-b b^*\varphi_2^\dagger \varphi_3$,\\
 $(\phi_1^\dagger \phi_2)^2=a^2\phi_1^\dagger\varphi_2 \phi_1^\dagger\varphi_2-bb^*\phi_1^\dagger\varphi_3 \phi_1^\dagger\varphi_3$,\\
$(\phi_1^\dagger \phi_3)^2=-bb^*\phi_1^\dagger\varphi_2 \phi_1^\dagger\varphi_2+a^2\phi_1^\dagger\varphi_3 \phi_1^\dagger\varphi_3$,\\
$(\phi_1^\dagger \phi_3)(\phi_1^\dagger \phi_2)=a^2\phi_1^\dagger\varphi_3 \phi_1^\dagger\varphi_2+b b^*\phi_1^\dagger\varphi_2^\dagger \phi_1^\dagger\varphi_3$\\
$(\phi_1^\dagger \phi_2)(\phi_2^\dagger \phi_1)=a^2\phi_1^\dagger\varphi_2 \varphi_2^\dagger \phi_1+bb^*\phi_1^\dagger\varphi_3 \varphi_3^\dagger\phi_1$,\\
$(\phi_1^\dagger \phi_3)(\phi_3^\dagger \phi_1)=bb^*\phi_1^\dagger\varphi_2 \varphi_2^\dagger\phi_1+a^2\phi_1^\dagger\varphi_3 \varphi_3^\dagger\phi_1$,\\
$(\phi_1^\dagger \phi_3)(\phi_2^\dagger \phi_1)=a^2\phi_1^\dagger\varphi_3 \varphi_2^\dagger\phi_1-b b^*\phi_1^\dagger\varphi_2 \varphi_3^\dagger\phi_1$

And also the complex conjugates. Note that $ab=0$. We can then define $a^2=\frac{1+\epsilon}{2}$ and $bb^*=\frac{1-\epsilon}{2}$. We can then combine the  $UU^*$-invariants into linearly independent polynomials which are either proportional to $\epsilon$ or absent from $\epsilon$. We can redefine $\epsilon(\theta)=ie^{i\theta}-ie^{-i\theta}$ with the phase  $\theta\to -\theta$ under a CP transformation (note that the imaginary unit commutes with CP, since it is the time-reversal which is anti-unitary). There is then a basis for the Higgs potential where the phase $\theta$ is absent if and only if the order 4 CP transformation is conserved. Note that this does not strictly invalidate the claim that a real basis does not exist~\cite{Ivanov:2015mwl}, because we are not using a complex notation and we are using background fields instead.

What it does show is that 
to deal with generalized CP it is better to treat the Higgs field as a real field and use a generalized basis involving a background field which transforms covariantly with the generalized CP transformation.
The standard bases (that do not involve background fields) transform under generalized CP as if it was a standard CP transformation, thus such bases are only good to handle standard CP transformations of the Higgs fields. Therefore we believe that the model of Ref.~\cite{Ivanov:2015mwl} gives support to our use of background fields in the CP basis.

 Note that our method requires knowledge of the $G_f$ group of family symmetries. There are alternative methods which do not require such knowledge~\cite{Varzielas:2016zjc}, but they are also complicated and we cannot guarantee that in a realistic situation it is not better to determine first which are the family symmetries $G_f$ (such knowledge is required for other purposes anyway).

A generalized CP transformation is different than a $Z_2$-like CP transformation only once we access the $G_f$-variant degrees of freedom.
In fact, the (background) CP symmetry of the Standard Model (without any extra degrees of freedom) may already be $Z_4$-like and we have no way to know it without new experimental results~\cite{pin}. For instance, if the neutrino is a Majorana particle, CP is order-4~\cite{pin,brancocp}. 

\subsection{Background symmetry in geometrical CP violation and rephasing symmetries}

In $n$-Higgs-doublet models, if $G_f/SU(2)_L$ is abelian and $G_f$ is a subgroup of $U(n)\times SU(2)_L$, then it commutes with the group of rephasing transformations of the Higgs fields $G_b=(U(1)^{n-1}\times U(1)_{Y}\times SU(2)_L)/Z_2$. Then any neutral Higgs field can be written as $\phi=e^{i\sum_{k=1}^{n-1} \theta_k}\phi_0$ with $\phi_0$ real and so verifying $c\phi_0=\phi_0$ where $c$ is the complex conjugation (related with the CP symmetry) and $\theta_k$ are phases parametrizing the group $G_b/(U(1)_{em}\times SU(2)_L)$.

Therefore, the background symmetry $G/G_b=Z_2$ cannot be explicitly broken by a source field which is a copy of a neutral Higgs field. It also cannot be spontaneously broken by a neutral vacuum expectation value of the Higgs field. In the particular case of the Standard Model, $U(1)^{n-1}$ is trivial and thus CP cannot be spontaneously broken.

Another application of background symmetries follows. Geometrical CP-violation involves calculable CP-violating phases~\cite{Branco:1983tn,Branco:2015hea,Fallbacher:2015rea,Holthausen:2012dk,*Varzielas:2012pd,*Ivanov:2013nla,realizable}. These phases are calculable in the sense that they are stable with respect to a finite variation of the parameters of the potential~\cite{Weinberg:1973am,Weinberg:1972ws}. The radiative corrections to calculable parameters are finite (and may be small but not necessarily) in a renormalizable model. The $\rho$ parameter is an example of a calculable parameter,
protected by the approximate custodial symmetry. The phase of the Higgs vev being calculable means that the infinite corrections to the Higgs vev have a fixed phase and thus the perturbative corrections to the phase of the Higgs vev come from the finite corrections to the Higgs vev.

The idea of spontaneous geometrical CP-violation arose in a three-Higgs-doublet model, with a $\Delta(54)$-symmetric Higgs potential which is a polynomial of fourth order. There is also explicit geometrical CP-violation~\cite{Branco:2015gna,Branco:2015hea}.

We describe it not as CP-violation, but as CP conservation up to a background phase. So we are dealing with CP as a background symmetry.

We consider a three-Higgs-doublet model, with explicit symmetry 
$G_f=(\Delta(54)/Z_3)\times U(1)_{Y}\times SU(2)_L$.
Promoting the parameters of the fourth order potential to background fields,
we have a background symmetry $G=G_b\rtimes Z_2$, with $G_b=(\Sigma(216\times 3)/Z_3)\times U(1)_{Y}\times SU(2)_L$,  
$G/G_f=A_4\rtimes Z_2\simeq S_4$~\cite{Grimus:2010ak,*Merle:2011vy,*Fairbairn:1964sga}.

Then, we choose as source field a copy of the Higgs field with the constraint $GJ=G_bJ$, thus  there is $h\in G_b$ such that $J=hJ_0$ with $J_0$ real (the CP acts as the complex conjugation here). Note that an arbitrary Higgs field does not verify $GJ=G_bJ$ for this case. Such constraint may be consequence of the minimization of a particular potential as in~\cite{Branco:1983tn} (see also Sec.~\ref{sec:pseudo}) or by the field content in the Action as in~\cite{Branco:2015gna,Branco:2015hea}.

Following~\cite{Fallbacher:2015rea}, we have the following doublet representations of $\Delta(54)/Z_3$ constructed from the three complex Higgs doublets $\phi_m$:
$(a_1,a_1^*)$,$(a_2,a_2^*)$,$(a_3,a_3^*)$ and (note the difference) $(a_4^*,a_4)$, where

\begin{align*}
\left[\begin{smallmatrix}
a_0\\
a_2\\
a_2^*
\end{smallmatrix}\right]=
M
\left[\begin{smallmatrix}
\phi_1^\dagger \phi_1\\
\phi_2^\dagger \phi_2\\
\phi_3^\dagger \phi_3
\end{smallmatrix}\right]=\frac{1}{\sqrt{3}}
\left[\begin{smallmatrix}
1 & 1 & 1\\
1 & \omega & \omega^2\\
1 & \omega^2 & \omega
\end{smallmatrix}\right]
\left[\begin{smallmatrix}
\phi_1^\dagger \phi_1\\
\phi_2^\dagger \phi_2\\
\phi_3^\dagger \phi_3
\end{smallmatrix}\right]
\end{align*}

\begin{align*}
\left[\begin{smallmatrix}
a_1\\
a_3\\
a_4
\end{smallmatrix}\right]=
M\left[\begin{smallmatrix}
\phi_3^\dagger \phi_2\\
\phi_1^\dagger \phi_3\\
\phi_2^\dagger \phi_1
\end{smallmatrix}\right]
\end{align*}

and $\omega$ is a complex number such that $\omega^3=1$ and $\omega+\omega^2=-1$. Note that the only invariant tensor of the gauge group $U(1)_{em}\times SU(2)_L$ is the kronecker delta with the indices of the complex $SU(2)_L$ doublet (and algebraic combinations of the kronecker delta).

The matrix $M$ is unitary. Also, 
$M^2=\left[\begin{smallmatrix}
1 & 0 & 0\\
0 & 0 & 1\\
0 & 1 & 0
\end{smallmatrix}\right]$, $M^4=1$. So, the 9 degrees of freedom in $a_n$ describe  fully and linearly the (at most) 9 degrees of freedom of any hermitian matrix\footnote{Note that for the particular case of $\phi^\dagger_j\phi_k$, there are at most 6 degrees of freedom plus 2 non-negative degrees of freedom.}, such as $\phi^\dagger_j\phi_k$. 

Now we can write $\phi(\omega,a)=\phi^*(\omega^2,a^*)$ so the $\phi(\omega,a)$ is invariant under the the complex conjugation (CP) $\phi\to \phi^*$ followed by the exchange of $a_n\to a_n^*$ (which comes from the exchange $\phi_2\to \phi_3$ ) followed by the exchange $\omega\to \omega^2$.

Imposing that $\phi_0$ is invariant under CP $\phi\to \phi^*$, then  $a_3=a_4^*$ and $a_1=a_1^*$. The $G/G_f=S_4$ group permutes $a_n$ (leaving $a_0$ invariant) and so we can have CP-violation of $h\phi_0$ due to a permutation of the $a_n$. Such CP-violation is such that there is a CP background symmetry which is conserved.

We can then use the source field $h\phi_0$ to modify the Higgs potential, adding an explicit symmetry breaking term.

\section{Conclusion}
\label{sec:conclusion}

Dealing with concepts which are not rigorously defined (in the mathematical sense) can have advantages with respect to an approach where every concept is rigorously defined~\cite{newmath}. In the context of Electroweak physics that is necessarily the case since a rigorously defined non-abelian gauge Quantum Field Theory does not exist yet. Therefore, assumptions play a key role. 

But after making the usual perturbative assumptions some problems are still very complicated. That is the case of building extensions of the Standard Model\footnote{Using extensions of the 
Standard Model is a practical way to produce predictions for experiments. But like statistical inference~\cite{pvalue}, (new) physics is not just about producing numbers. E.g. accounting all reasonable extensions, we may have one prediction for each logical possibility~\cite{Schucker:2007ge}, which is a kind of look-elsewhere effect.
Producing predictions where such effect is consistently accounted for is a hard problem (even if we assume spontaneous symmetry breaking of $SU(2)_L$).}, and in particular studying the Higgs potential (a symmetric polynomial of many variables~\cite{Sartori:2003ac,*Sartori:1992ib,*Abud:1983id,stratifiedmorse,*stratifiedmorse2,Wybourne:1980eh,*O'Raifeartaigh:1986vq,*indefinite}).

We should be careful: making assumptions can be used to focus on the physical questions as much as it can be used to avoid the physical questions.

To study the Higgs potential, one option is to check what are the implications of alternative assumptions. Such as non-perturbative assumptions---e.g. the ones used in lattice gauge theory or in the functional renormalization group, which can produce complementary results~\cite{DeGrand:2015zxa,Eichhorn:2015kea}. Or working with real representations of groups---which in a real polynomial makes sense~\cite{realoperatoralgebras} and it is necessary\footnote{also to study the physical spectrum in multi-Higgs-doulet models; to handle the pseudo-goldstone bosons in multi-Higgs models; or to do lattice simulations of the Higgs sector.}  to deal with the approximated custodial symmetry of the Higgs potential~\cite{pdg,Pich:2015lkh,accidental2,accidental,Pilaftsis:2016erj,Nishi:2011gc}. 

In this paper we showed that such option does lead to progress, despite that the perturbative Electroweak expansion is a good approximation to the experimental results. We discussed several examples how the research on Grand-Unified-Theories and CP violation can be much improved by 
defining the spontaneous  breaking of a global symmetry  as a particular case of explicit symmetry breaking. Moreover, in this way such research can be complemented by non-perturbative studies.

In conclusion, assuming gauge symmetry breaking or using only complex representations of groups is not sufficient to study the phenomenology of extended Higgs sectors. 

\paragraph{Acknowledgments}
L.\ P.\ acknowledges the hospitality of the Institute of Physics at the University of Graz, where most of this work has been done, and of the
Centro de F\'isica Te\'orica de Part\'iculas at the Universidade de Lisboa.
L.P. acknowledges Axel Maas, Gustavo Branco, Igor Ivanov and Ivo Varzielas for useful conversations.

\appendix

\section{Adding explicit symmetry breaking terms to the Higgs potential}
\label{sec:proc}

Now we show that it is always possible to modify the tree-level potential by adding an infinitesimal term  such that its minimum is the one we want. Note that since there is a cutoff scale,   the radiative corrections due to the presence of the added infinitesimal term are infinitesimal as well, even when divergencies appear in these corrections. Thus the presence of the infinitesimal term
does not affect the non-infinitesimal part of the effective potential, even when radiative corrections are taken into account.

We consider a $G/G_b$-invariant and $G_f$-invariant Higgs potential of arbitrary order (i.e. eventually non-renormalizable). The subgroups $G_b$, $G_f\subset G_b$ (with the gauge group contained in $G_f$) are normal subgroups of $G$ which is a compact group.

We take the numerical values of the source field $J=\epsilon \phi_0$ proportional to the numerical values of a Higgs field $\phi_0$ with $N$ real components corresponding to the absolute minimum of the Higgs potential.

The important point is that the orthogonal group $O(N)$ ($G$ is compact) acts on the $\mathbb{R}^{\otimes n}$ tensor space with unitary operators.
There is a one-to-one correspondence between isomorphisms of a vector space 
$V$ to all of its dual space $V^*$ and nondegenerate bilinear forms on $V$.
Since $O(N)$ is compact, we can find a scalar product on V that makes the representation unitary.

For $n=1$, we have the basis $e_j$ and the bilinear form $<e_j,e_k>=\delta_{jk}$.
Such form makes the representation unitary.
For arbitrary $n$, we have the basis $e_{j1}\otimes...e_{jn}$ and the 2n-linear form 
$<e_{j1}\otimes...e_{jn},e_{k1}\otimes...e_{kn}>=\frac{1}{n!}\sum_\sigma \delta_{j1\sigma(k1)}...\delta_{jn\sigma(kn)}$.
Such form makes the representation of $O(n)$ unitary.

Therefore, suppose we have a polynomial strictly of order 2 in $\phi$. 
It can be written using the inner product of the 
tensors $p^{jk} e_j\otimes e_k$ and  $\phi^{j}\phi^k e_j\otimes e_k$

Now we want to find a $G_f$-invariant tensor whose maximum occurs at $\phi=\phi_0$, with $G_f\phi_0$ breaking $c\in G$.

There is a basis of $G_f$-invariant tensors, which is given by $\rho^{jk}_a e_j\otimes e_k$. The $\rho_a$ are chosen such that $<\rho_a,\rho_b>=\delta_{ab}$, using the Gram-Schmidt process. Note that the basis  $\rho_a$ is complete in the space of  $G_f$-invariant 2nd order tensors, but it is incomplete in the space of 2nd order tensors. We can complete it nevertheless. The first $n_f$ components are $G_f$-invariant, the remaining $n^2-n_f$ components are not $G_f$-invariant.
  
Then, $\rho^{lm}_a\rho^{jk}_a=\frac{\delta_{l}^j\delta_{m}^k+\delta_{l}^k\delta_{m}^j}{2}$ with $a$ running until $n^2$. 
We then write $\Phi^a=<\rho_a,\phi_0\otimes \phi_0>$ and $\Phi_f=\sum_{a=1}^{n_f}\Phi^a\rho_a$ is a $G_f$-invariant tensor. 

Then, we can write any tensor as $T=c R\Phi_f$ where $c>0$ is a normalization factor and $R$ is a $O(n^2)$ rotation.
Then, the $R$ which maximizes the inner-product $<T,\Phi_f>$ is $R=1$.
Of course, not all $R$ is meaningful. $R$ should be such that it can be written as a $O(n)\otimes O(n)$ rotation.
But since the representation of $O(n)$ is unitary, then $O(n)\otimes O(n)\subset O(n^2)$ and so $R=1\in O(n)\otimes O(n)$ is a valid rotation.
Also, $T=c\Phi_f$ is a $G_f$-invariant tensor which can be written as $T=\sum_{a=1}^{n_K}<\rho_a,\phi\otimes \phi>\rho_a$.

Therefore, the Higgs field minimizing the polynomial of order 2 
$V_2=-\sum_{a=1}^{n_K}<\rho_a,\phi\otimes \phi><\rho_a,\Phi_f>$ is $\phi_0$. 
At each order we can do the same and so we conclude that we can always add an infinitesimal $G_f$-invariant polynomial (defined by a source field $J=\epsilon \phi_0$) such that the chosen $G_f$-orbit minimizing the modified potential is not related by $c\in G$ to another absolute minimum.

Note that since the potential $V$ has a background symmetry $G$ then $V_B(\phi)=V_{c B}(c \phi)$ and also the reference point transforms under $G$ covariantly with respect to the background fields~\cite{Fallbacher:2015rea}, i.e. 
$W_{B,J}(\phi)=W_{g B, gJ}(g \phi)$ for all $g\in G$. By construction, a background transformation $c\in G$ is conserved  (i.e. it is not spontaneously broken) if and only if $c$ is conserved by the modified Higgs potential 
$W_{B,J}(\phi)=W_{c h B,J}(c h \phi)$ for some $h\in G_b$.
Therefore, the numerical values of background fields and source fields, suffice to determine the spontaneously broken symmetries.

\addcontentsline{toc}{section}{References}
\footnotesize
\singlespacing 
\bibliography{Poincare}
\bibliographystyle{utphysMM}
\begin{acronym}[nuMSM]
\acro{2HDM}{two-Higgs-doublet model}
\acro{ATLAS}{A Toroidal LHC ApparatuS}
\acro{BR}{Branching Ratio}
\acro{BGL}{Branco\textendash{}Grimus\textendash{}Lavoura}
\acro{BSM}{Beyond the Standard Model}
\acro{CL}{Confidence Level}
\acro{cLFV}{charged Lepton Flavor Violation}
\acro{CLIC}{Compact Linear Collider}
\acro{CMS}{Compact Muon Solenoid}
\acro{CP}{Charge-Parity}
\acro{CPT}{Charge-Parity-Time reversal}
\acro{DM}{Darkmatter}
\acro{EDM}{Electric Dipole Moment}
\acro{EFT}{Effective Field Theory}
\acro{EW}{Electroweak}
\acro{EWSB}{Electroweak symmetry breaking}
\acro{FCNC}{Flavour Changing Neutral Current}
\acro{MET}{Missing Transverse Energy}
\acro{MFV2}{Minimal Flavor Violation with two spurions}
\acro{MFV6}{Minimal Flavor Violation with six spurions}
\acro{GIM}{Glashow\textendash{}Iliopoulos\textendash{}Maiani}
\acro{GNS}{Gelfand-Naimark-Segal}
\acro{GUT}{Grand unified theory}
\acro{ILC}{International linear collider}
\acro{LEP}{Large electron\textendash{}positron collider}
\acro{LFC}{Lepton flavor conservation}
\acro{LFV}{Lepton Flavor Violation}
\acro{LHC}{Large Hadron Collider}
\acro{MFV}{Minimal flavour violation}
\acro{MIA}{Mass insertion approximation}
\acro{MSSM}{Minimal Supersymmetry Standard Model}
\acro{nuMSM}[$\nu$MSM]{minimal extension of the Standard Model by three right-handed neutrinos}
\acro{PS}{Pati-Salam}
\acro{PT}[$\mathrm{p_T}$]{transverse momentum}
\acro{QCD}{Quantum chromodynamics}
\acro{RG}{Renormalization group}
\acro{RGE}{Renormalization group equation}
\acro{SM}{Standard Model}
\acro{SUSY}{Supersymmetry, Supersymmetric}
\acro{VEV}{Vacuum expectation value}
\acro{MEG}{Muon to electron and gamma}
\acro{NP}{New Physics}
\acro{NH}{Normal hierarchy}
\acro{IH}{Inverted hierarchy}
\acro{CKM}{Cabibbo\textendash{}Kobayashi\textendash{}Maskawa}
\acro{PMNS}{Pontecorvo-Maki-Nakagawa-Sakata}

\end{acronym}
\normalsize
\onehalfspacing 
\end{document}